# Explicit constructions of all separable two-qubits density matrices and related problems for three-qubits systems

Y. Ben-Aryeh and A. Mann

*Physics Department, Technion-Israel Institute of Technology, Haifa 32000, Israel*

e-mails: phr65yb@physics.technion.ac.il ; ady@physics.technion.ac.il

ABSTRACT

Explicitly separable density matrices are constructed for all separable two-qubits states based on Hilbert-Schmidt (HS) decompositions. For density matrices which include only two-qubits correlations the number of HS parameters is reduced to 3 by using local rotations, and for two-qubits states which include single qubit measurements, the number of parameters is reduced to 4 by local Lorentz transformations. For both cases we related the absolute values of the HS parameters to probabilities, and the outer products of various Pauli matrices were transformed to pure state density matrices products. We discuss related problems for three qubits. For n-qubits correlation systems ($n \geq 2$) the sufficient condition for separability may be improved by local transformations, related to high order singular value decompositions.

1. Introduction

For systems with many subsystems, and Hilbert spaces of large dimensions, the "separability problem" becomes quite complicated.[1] In the simple case of two-qubits states, it is possible to give a measure of the degree of quantum correlations by using the partial-transpose (PT) of the density matrix.[1-3] According to the Peres-Horodecki (P-H) criterion[2-3] if the PT of the two qubits density matrix leads to negative eigenvalues of the PT matrix $\rho_{AB}(PT)$, then the density matrix is entangled, otherwise it is separable.

One should take into account that the density operator of a given mixture of quantum states has many ensemble decompositions. The separability problem for two-qubits states is defined as follows: A bipartite system is separable if and only if the density matrix of this system can be written in the form:

$$\rho = \sum_j p_j \rho_A^{(j)} \otimes \rho_B^{(j)} \quad . \tag{1}$$

Here: $p_j > 0$ and $\sum_j p_j = 1$. The density matrix $\rho_{AB}$ is defined on the Hilbert space $H_A \otimes H_B$ where $A$ and $B$ are the two parts of the bipartite system. $\rho_A^{(j)}$ and $\rho_B^{(j)}$ are density matrices for the A and B systems, respectively. The interpretation of such definition is that for bipartite separable states the two systems are independent of each other. The summation over $j$ could include large numbers of density matrices products, but usually it is preferred to limit this number to smaller ones, as far as it is possible.

The density matrix for any bipartite system can be given as

$$\rho_{AB} = \sum_{i,k,j,l} \rho_{ik,jl} \left(|i\rangle\langle j|\right)_A \otimes \left(|k\rangle\langle l|\right)_B \quad . \tag{2}$$

Here $|i\rangle$ and $\langle j|$ are summed over a complete set of states of the $A$ system, while $|k\rangle$ and $\langle l|$ states are summed over the $B$ system. $\rho_{ik,jl}$ are complex numbers. The PT of the density matrix $\rho_{AB}$ is given as:

$$\rho_{AB}(PT) = \sum_{i,l,j,k} \rho_{il,jk} \left(|i\rangle\langle j|\right)_A \otimes \left(|k\rangle\langle l|\right)_B \quad . \tag{3}$$

M. Horodecki, et al. [3] showed that the positivity of the PT state is necessary and sufficient for separability of $2\times 2$ and $2\times 3$ systems. However, explicit expressions for separable density matrices of two qubits have not been presented. One aim of the present work is to exhibit explicitly a separable form for any separable 2-qubits state, and discuss the possibilities to use the present methods for more qubits.

For a 2-qubits system denoted by A and B we use the Hilbert-Schmidt (HS) representation of the density matrix: [4,5]

$$\rho_{AB} = \tfrac{1}{4}\left\{(I)_A \otimes (I)_B + (\vec{r}\cdot\vec{\sigma})_A \otimes (I)_B + (I)_A \otimes (\vec{s}\cdot\vec{\sigma})_B + \sum_{m,n=1}^{3} t_{m,n}(\sigma_m)_A \otimes (\sigma_n)_B\right\} . \tag{4}$$

Here, $I$ represents the unit $2\times 2$ matrix, the three Pauli matrices are denoted by $\vec{\sigma}$ while $\vec{r}$ and $\vec{s}$ represent 3-dimensional parameters–vectors. The normalized 2-qubits density matrix is described by 15 real HS-parameters: 6 for $\vec{r}$ and $\vec{s}$, and 9 for $t_{m,n}$. The quantum correlations are included in the HS $t_{m,n}$ parameters while $\vec{r}$ and $\vec{s}$ correspond to single qubit measurements. Assuming the conditions $\vec{r} = \vec{s} = 0$ in (4), we get the density matrix

$$4\rho_{AB} = \left\{(I)_A \otimes (I)_B + \sum_{m,n=1}^{3} t_{m,n}(\sigma_m)_A \otimes (\sigma_n)_B\right\} . \tag{5}$$

We show in the present work a simple method by which the HS decompositions (4) and (5) can be transformed to the form of (1) for all two-qubits density matrices which are separable. The analysis for separability for the density matrices (4) and (5) becomes in the general case quite complicated due to the large number of parameters involved in such analysis (15 for (4), and (9) for (5)). However, the number of parameters describing the 2-qubits density matrices can be reduced by local transformations.[6–11] We consider $\rho$ and $\rho_M$ to be of the same equivalence class when

$$\rho \to \rho^M = M\rho M^\dagger \quad , \quad M = M_A \otimes M_B \quad , \tag{6}$$

where $M_A$ and $M_B$ are invertible. Such equivalence preserves the positivity and separability of the density matrices. For density matrices of the form (5), $M_A$ and $M_B$ are given by orthogonal rotation matrices so that normalization is preserved. For density matrices of the form (4), $M_A$ and $M_B$ are given by Lorentz transformations,[7] so the transformed density matrix should be renormalized. These local transformations can reduce the number of parameters for the density matrix (5) to 3 parameters by using rotations, and the number of parameters for the density matrix (4) to 4 parameters, by local Lorentz transformations.[7] While such transformations have been analyzed,[7] the explicitly separable density matrices for 2-qubits density matrices have not been analyzed and such explicitly separable density matrices are treated in the present work. We derive simple necessary and sufficient conditions for separability which are in agreement with the results given by other authors[12–14] but obtained here by using a different method. We discuss the use of the present method for larger n-qubits systems (n>2).

The present paper is arranged as follows:

In Sec. 2 we describe explicitly separable density matrices of 2-qubits density matrices of the form (5), where the number of parameters is reduced by local rotations. In Sec. (3) we relate the analysis of separable density matrices of 2-qubits states of the form (4), by the HS decompositions, to an analysis made previously by local Lorentz transformations[7] and show the explicit separable form in the generic[11] case. In Sec. (4) we discuss certain problems related to the application of the present methods to larger n-qubits correlation systems ($n > 2$). In Sec. (5) we summarize our results.

## 2. An explicit construction of separable density matrices of 2-qubits states described by the HS decomposition of Eq. (5)

One may use local rotations to bring the form $\sum_{m,n=1}^{3} t_{m,n} (\sigma_m)_A \otimes (\sigma_n)_B$ into diagonal form: $\sum_{i=1}^{3} t_i (\sigma_i)_A \otimes (\sigma_i)_B$, as follows. Since the $\sigma_i$ $(i=1,2,3)$ transform as a vector under rotations, we may define:

$$(\sigma_m)_A = (O_{m,i})_A (\bar{\sigma}_i)_A \quad , \quad (\sigma_n)_B = (O_{n,j})_B (\bar{\sigma}_j)_B \quad , \tag{7}$$

where $(O)_A, (O)_B$ are 3×3 orthogonal matrices. Hence

$$\sum_{m,n=1}^{3} t_{m,n} (\sigma_m)_A \otimes (\sigma_n)_B = \sum_{m,n,i,j} (O_{m,i})_A t_{m,n} (O_{n,j})_B (\bar{\sigma}_i)_A \otimes (\bar{\sigma}_j)_B \quad . \tag{8}$$

By the singular-value decomposition (SVD)[15,16] we can choose $(O)_A$ and $(O)_B$ so that

$$\sum_{m,n} (O_{m,i})_A t_{m,n} (O_{n,j})_B = \delta_{ij} t_i \quad , \tag{9}$$

and therefore

$$\sum_{m,n=1}^{3} t_{m,n} (\sigma_m)_A \otimes (\sigma_n)_B = \sum_{i=1}^{3} t_i (\bar{\sigma}_i)_A \otimes (\bar{\sigma}_i)_B \quad . \tag{10}$$

We discard the bars in (10) and analyze the construction of separable density matrices for the density matrix given as:

$$4\rho_{AB} = \left\{ (I)_A \otimes (I)_B + \sum_{i=1}^{3} t_i (\sigma_i)_A \otimes (\sigma_i)_B \right\} \quad . \tag{11}$$

It is easy to verify that $\rho_{AB}$ of Eq. (11) may be written as

$$4\rho_{AB} =$$

$$\sum_{i=1}^{3} 2|t_i| \left[ \left\{ \frac{(I-\sigma_i)_A}{2} \otimes \frac{(I-sign(t_i)\sigma_i)_B}{2} \right\} + \left\{ \frac{(I+\sigma_i)_A}{2} \otimes \frac{(I+sign(t_i)\sigma_i)_B}{2} \right\} \right] \quad (12)$$

$$+ \left[ (I)_A \otimes (I)_B \right] \left( 1 - \sum_{i=1}^{3} |t_i| \right)$$

Here for positive $t_i$ $sign(t_i)=1$, and for negative $t_i$ $sign(t_i)=-1$. Each outer product in the curly brackets on the right side of (12) is a separable density matrix, based on pure states. $|t_i|/2$ can be considered as a probability for each pure state separable density matrix included in the curly brackets. The unit matrix $\left[ (I)_A \otimes (I)_B \right]$ (in the last term of (12)) is trivially a separable density matrix. Therefore, if its coefficient $\left( 1 - \sum_{i=1}^{3} |t_i| \right)$ is nonnegative, then $\rho_{AB}$ is separable.

Hence $\sum_{i=1}^{3} |t_i| \leq 1$ is a *sufficient* condition for separability. We will prove that this condition is also a *necessary* condition by using a different method from that presented in previous works. Our idea is that the present methods can be generalized to larger n-qubits systems ($n > 2$)

To prove that the above condition is *necessary* assume that $\rho_{AB}$ is separable. As pointed out by Peres,[2] if $\rho_{AB}$ is separable then $\rho_{AB}(PT)$ is also a separable density matrix. In the standard basis $|00\rangle, |01\rangle, |10\rangle, |11\rangle$, the density matrix (11, 12) is given by:

$$4\rho_{AB} = \begin{pmatrix} 1+t_3 & 0 & 0 & t_1-t_2 \\ 0 & 1-t_3 & t_1+t_2 & 0 \\ 0 & t_1+t_2 & 1-t_3 & 0 \\ t_1-t_2 & 0 & 0 & 1+t_3 \end{pmatrix} \quad (13)$$

The PT of the density matrix (13) is given by inverting the sign of $t_2$,[4] so

$$4\rho_{AB}(PT) = \begin{pmatrix} 1+t_3 & 0 & 0 & t_1+t_2 \\ 0 & 1-t_3 & t_1-t_2 & 0 \\ 0 & t_1-t_2 & 1-t_3 & 0 \\ t_1+t_2 & 0 & 0 & 1+t_3 \end{pmatrix}. \quad (14)$$

The eigenvalues of $\rho_{AB}(PT)$ are given by:

$$\begin{aligned} 4\lambda_1(PT) &= 1+t_1-t_2-t_3, & 4\lambda_2(PT) &= 1-t_1+t_2-t_3, \\ 4\lambda_3(PT) &= 1-t_1-t_2+t_3, & 4\lambda_4(PT) &= 1+t_1+t_2+t_3 \end{aligned}. \quad (15)$$

The eigenvalues of $\rho_{AB}$ are given by

$$\begin{aligned} 4\lambda_1(\rho) &= 1-t_1-t_2-t_3, & 4\lambda_2(\rho) &= 1+t_1+t_2-t_3, \\ 4\lambda_3(\rho) &= 1+t_1-t_2+t_3, & 4\lambda_4(\rho) &= 1-t_1+t_2+t_3 \end{aligned}. \quad (16)$$

We note that in (15, 16) inverting the signs of two $t_i$ merely exchanges the names of the eigenvalues. We note also that separability of $\rho_{AB}$ ensures that all these eight eigenvalues are nonnegative. One should notice also that by inverting the sign of any one of the $t_i$ parameters the eigenvalues of $\rho_{AB}(PT)$ are transformed into the eigenvalues of $\rho_{AB}$, and vice versa. The important point is that one has to look at both the eigenvalues of $\rho_{AB}$ and $\rho_{AB}(PT)$.

The necessity of the condition

$$\sum_{i=1}^{3} |t_i| \leq 1, \quad (17)$$

have been derived by other authors,[12–14] but we derive it by using a different method. Indeed it is easy to verify that whatever the sign of $t_i (i=1,2,3)$, $1-\sum_{i=1}^{3}|t_i|$ is always one of the 8 eigenvalues given in (15,16) and therefore non-negative if $\rho$ is separable. If the sign of $t_1 t_2 t_3$ is positive, we see that $1-\sum_{i=1}^{3}|t_i|$ is an eigenvalue of $\rho$, and therefore nonnegative and $\rho$ is separable. If the sign of $t_1 t_2 t_3$ is negative, then $1 - \sum_{i=1}^{3} |t_i|$ is an eigenvalue of $\rho(PT)$ and therefore nonnegative if $\rho$ is separable.

Further interesting relations and even a simpler necessary and sufficient condition for separability have been derived previously[12] and may be obtained here in the following way. We can invert the relations given by (16) so that the HS parameters $t_i$ ($i = 1, 2, 3$) will be given as functions of the eigenvalues. Then we get:

$$t_1 = 1 - 2\lambda_2(\rho) - 2\lambda_3(\rho) \quad ; \quad t_2 = 1 - 2\lambda_2(\rho) - 2\lambda_4(\rho) \quad ; \quad t_3 = 2\lambda_3(\rho) + 2\lambda_4(\rho) - 1 \ . \quad (18)$$

Substituting these values in (15) we get:

$$\lambda_1(PT) = (1/2) - \lambda_3(\rho) \quad ; \quad \lambda_2(PT) = (1/2) - \lambda_4(\rho) \quad ;$$
$$\lambda_3(PT) = (1/2) - \lambda_1(\rho) \quad ; \quad \lambda_3(PT) = (1/2) - \lambda_2(\rho) \quad . \quad (19)$$

We find that if all eigenvalues of $\rho_{AB}$ are less than or equal $1/2$, then $\rho_{AB}$ is separable. But if one of the eigenvalues of $\rho_{AB}$ is larger than $1/2$ then $\rho_{AB}$ is entangled.[12]

**3. Explicitly separable density matrices of 2-qubits states described by the general case of HS decomposition given by Eq. (4)**

Let us assume a two-qubits density matrix in which the matrix $t_{mn}$ has been diagonalized but it includes also single qubits elements. This density matrix is given as:

$$\rho_{AB} = \tfrac{1}{4}\left\{ (I)_A \otimes (I)_B + (\vec{a} \cdot \vec{\sigma})_A \otimes (I)_B + (I)_A \otimes (\vec{b} \cdot \vec{\sigma})_B + \sum_{i=1}^{3} t_i (\sigma_i)_A \otimes (\sigma_i)_B \right\} \quad , \quad (20)$$

where $\vec{a}$ and $\vec{b}$ are certain parameters-vectors satisfying the relations $|\vec{a}| \leq 1$ and $|\vec{b}| \leq 1$. This density matrix can be written as:

$$4\rho_{AB} = \sum_{i=1}^{3} \frac{|t_i|}{2} \left[ \left\{ \frac{I + \text{sign}(t_i)\sigma_i}{2} \right\}_A \otimes \left\{ \frac{I + \sigma_i}{2} \right\}_B + \left\{ \frac{I - \text{sign}(t_i)\sigma_i}{2} \right\}_A \otimes \left\{ \frac{I - \sigma_i}{2} \right\}_B \right] +$$

$$a\left[\left(I + \frac{\vec{a}\cdot\vec{\sigma}}{a}\right)/2\right]_A \otimes \left(\frac{I}{2}\right)_B + b\left(\frac{I}{2}\right)_A \otimes \left[\left(I + \frac{\vec{b}\cdot\vec{\sigma}}{b}\right)/2\right]_B + (I)_A \otimes (I)_B \left(1 - a - b - \sum_{i=1}^{3} |t_i|\right)$$

, (21)

where $a = |\vec{a}|, b = |\vec{b}|$. A sufficient condition for separability is given by

$$\left(1 - a - b - \sum_{i=1}^{3} |t_i|\right) \geq 0 .  \tag{22}$$

However, it is not necessary unless $|\vec{a}| = |\vec{b}| = 0$.

To obtain a necessary and sufficient condition and explicitly separable form for the density matrix we note that the general density matrix (4) can be written in the HS representation as [6–11]

$$4\rho = \sum_{\alpha,\beta=0}^{3} R_{\alpha,\beta} (\sigma_\alpha)_A \otimes (\sigma_\beta)_B . \tag{23}$$

Here the summation extends from 0 to 3, $\sigma_0$ is the $2\times 2$ identity matrix, and $\sigma_1, \sigma_2, \sigma_3$ are the Pauli spin matrices. Verstraete et al.,[7,11] studied how the density matrix is changed under local quantum operations and classical communications (LQCC) of the type

$$\rho' \rightarrow (A \otimes B) \rho (A \otimes B)^\dagger . \tag{24}$$

They[7,11] have shown that the $4\times 4$ matrix $R_{\alpha\beta}$ can be given as

$$R = L_1 \Sigma L_2^T . \tag{25}$$

Here $L_1$ and $L_2$ are proper local Lorentz transformations, and $\Sigma$ is either the diagonal form $\Sigma = diag(s_0, s_1, s_2, s_3)$ (the generic case[11] with $s_0 \geq s_1 \geq s_2 \geq |s_3|$ where $s_3$ is positive or negative) or of the form

$$\Sigma = \begin{pmatrix} a & 0 & 0 & b \\ 0 & d & 0 & 0 \\ 0 & 0 & -d & 0 \\ c & 0 & 0 & a+c-b \end{pmatrix} . \qquad (26)$$

Here $a,b,c,d$ are real parameters satisfying one of the four relations:[11]

a. For $b=c=a/2$, $\rho_{AB}$ is always entangled.[11]

b. For the other 3 cases, $\rho_{AB}$ is separable and has a simple form given explicitly by them.[11]

More detailed analysis for the density matrix (23) has been given by Caban et al.[17]

For the generic case, diagonal $\Sigma$,[11] the situation is very similar to that which we analyzed in Section 2. (Note that in Section 2 we could have chosen $1 \geq t_1 \geq t_2 \geq |t_3|$ ). Therefore in this case

$$\rho_{AB} = \sum_{i=1}^{2} \frac{s_i}{2} \left[ \left\{ \frac{(I+\sigma_i)_A}{2} \otimes \frac{(I+\sigma_i)_B}{2} \right\} + \left\{ \frac{(I-\sigma_i)_A}{2} \otimes \frac{(I-\sigma_i)_B}{2} \right\} \right]$$
$$+ \frac{|s_3|}{2} \left[ \left\{ \frac{(I+\sigma_3)_A}{2} \otimes \frac{(I+\mathrm{sign}(s_3)\sigma_3)_B}{2} \right\} + \left\{ \frac{(I-\sigma_3)_A}{2} \otimes \frac{(I-\mathrm{sign}(s_3)\sigma_3)_B}{2} \right\} \right] \qquad (27)$$
$$+ \left[ \frac{(I)_A \otimes (I)_B}{4} \right] \left[ s_0 - (s_1 + s_2 + |s_3|) \right]$$

The necessary and sufficient condition for separability is

$$s_1 + s_2 + |s_3| \leq s_0 \quad . \qquad (28)$$

Also this condition may again be written in terms of the eigenvalues of $\rho_{AB}$ as

$$\lambda_{AB}^{i} \leq s_0/2 \quad (i=1,2,3,4) \quad . \qquad (29)$$

### 4. Sufficient conditions for separability and entanglement for correlated systems of 3-qubits

A three-partite system is separable if and only if the density matrix of this system can be written in the form:

$$\rho = \sum_j p_j \rho_A^{(j)} \otimes \rho_B^{(j)} \otimes \rho_C^{(j)} \quad . \tag{30}$$

Here: $p_j > 0$ and $\sum_j p_j = 1$.

We are interested in genuine 3 qubits correlations density matrices like that of the two-qubits correlations given by the density matrix (5). We assume a 3 qubits correlation density matrix given by:

$$8\rho_{ABC} = (I)_A \otimes (I)_B \otimes (I)_B + \sum_{a,b,c=1}^{3} G_{a,b,c} (\sigma_a)_A \otimes (\sigma_b)_B \otimes (\sigma_c)_C \quad . \tag{31}$$

We find that the GHZ density matrices[18] and braid density matrices[19] are special forms of (31). These examples are, however, based on pure states while the present analysis for the sufficient conditions for separability will be more general, as we include in (31) mixed density matrices. We study the relations between the real HS parameters $G_{a,b,c}$, which can be either positive or negative, with the sufficient conditions for separability of (31).

A separable-like form for the density matrix (31) can be given as:

$$8\rho_{ABC} =$$

$$(1/4) \sum_{a,b,c=1}^{3} |G_{a,b,c}| \cdot \begin{cases} \left[\{(I)_A+(\sigma_a)_A\} \otimes \{(I)_B-(\sigma_b)_B\} \otimes \{(I)_C-sign(G_{abc})(\sigma_c)_C\}\right]+ \\ \left[\{(I)_A+(\sigma_a)_A\} \otimes \{(I)_B+(\sigma_b)_B\} \otimes \{(I)_C+sign(G_{abc})(\sigma_c)_C\}\right]+ \\ \left[\{(I)_A-(\sigma_a)_A\} \otimes \{(I)_B-(\sigma_b)_B\} \otimes \{(I)_C+sign(G_{abc})(\sigma_c)_C\}\right]+ \\ \left[\{(I)_A-(\sigma_\alpha)_A\} \otimes \{(I)_B+(\sigma_\beta)_B\} \otimes \{(I)_C-sign(G_{abc})(\sigma_c)_C\}\right] \end{cases}, \tag{32}$$

$$+ \left(1 - \sum_{a,b,c=1}^{3} |G_{a,b,c}|\right) \{(I)_A\} \otimes \{(I)_B\} \otimes \{(I)_c\}$$

Each expression in the curly brackets of (32) represents a pure state density matrix multiplied by 2. We get according to (32) that a sufficient condition for separability is given by

$$\left(\sum_{a,b,c=1}^{3}|G_{a,b,c}|\right)\leq 1 \quad \textit{sufficient condition for separability} \quad . \tag{33}$$

The crucial point here is that the form $\left(\sum_{a,b,c=1}^{3}|G_{a,b,c}|\right)$ is not invariant under orthogonal transformations and we can improve the condition for separabilty by using orthogonal transformations which will minimize this "separability form". For the density matrix (31) we can use for each qubit a rotation which will include 3 parameters. For the general mixed 3-qubits correlation systems we have 27 parameters, but the rotations of the 3-qubits system will introduce only 9 parameters. So we are looking for minimizing the function $\sum_{a,b,c=1}^{3}|G_{a,b,c}|$ by varying 9 rotation parameters, which is a very complicated problem related to high order singular value decompositions (HOSVD). The situation is quite different for the case of two qubits where the corresponding "separability form" is given by $\sum_{m,n=1}^{3}|t_{m,n}|$. We get 9 parameters and a sufficient condition for separability is given by $\sum_{m,n=1}^{3}|t_{mn}|\leq 1$. For two qubits there are 6 rotations parameters, and by varying these parameters we reduced it to three diagonal parameters $t_i$ $(i=1,2,3)$, where we obtain Eq. (17) as sufficient and necessary condition for separability.

The P-H criterion, that $\rho(PT)$ has a negative eigenvalue, is sufficient for entanglement. Eq. (33) gives a sufficient condition for separability. Hence if the eigenvalues of $\rho(PT)$ are positive and therefore the P-H criterion does not work, Eq. (33) may be tried as a sufficient condition for separability. Of course, usually there will be a region where both criteria will not help.

The present approach can be generalized to larger correlation n-qubits systems (n>3) by minimizing the separability form using HOSVD. One can use also numerical analysis for minimizing the separability form.

## 5. Summary

In the present article we have exhibited an explicitly separable form for all separable two qubits density matrices (Eqs. (12), (27)). The present method for deriving the necessary and sufficient conditions for two-qubits correlation states (Eqs. (17) and (19)) is different from those given in other works and can be generalized to larger n-qubits systems (n>2). For 3-qubits correlation system we have exhibited a possible separable form (Eq. (32)) for all density matrices satisfying a sufficient condition for separability (Eq. (33)). We discussed orthogonal transformations which can improve the condition for separability.